\documentclass[aps,pre,onecolumn,showpacs,superscriptaddress,nofootinbib]{revtex4-1}

\usepackage[pdftex]{graphicx}

\usepackage{amsmath}
\usepackage[latin1]{inputenc}
\usepackage{txfonts}
\usepackage{hyperref}

\usepackage{xcolor}

\begin{document}

\title{Approximate solution for the oscillation death state in pulse coupled oscillators}
\author{Rafael S. Pinto}
\email{rsoaresp@gmail.com}
\affiliation{Institute of Mathematical and Computer Sciences, Universidade de São Paulo, São Carlos 13566-590, São Paulo, Brazil}

\date{\today}

\begin{abstract}
We study numerically the oscillation death state in the phase oscillator model proposed by Winfree. We found that the phases in this state follow very simple rules, actually, besides intrinsic properties of the oscillators, such as natural frequency and pulse shape, they depend only on the inverse of the degree. Other topological properties such as transitivity or associativity seems to play no role on this state. Furthermore, we found that degree-frequency correlation helps to inhibit oscillation death. Simple analytical approximations corroborate the numerical results.
\end{abstract}

\maketitle

\section{Introduction}

The mathematical analysis of synchronization phenomena was probably shaped in its modern form from the early works of Winfree \cite{winfree1967,winfree1980}, who analyzing models of coupled limit cycles oscillators discovered that a mutual rhythm could emerge from a population of heterogeneous elements if the coupling strength between oscillators would surpass a critical value. Some years later, inspired by the works of Winfree \cite{strogatz2000}, Kuramoto achieved significant advances in the weak coupling limit, culminating in what is know today as the Kuramoto model \cite{kuramoto1975,strogatz2000,acebron2005,rodrigues2016}, that is almost synonym of synchronization.

The Kuramoto model is easier to handle mathematically than Winfree's model, as it's amenable to a self consistent approach that allows one to obtain analytically the threshold value of the coupling, as well as the behavior of the order parameter for some situations. Probably due to this, the Winfree model was eclipsed and attracted much less attention over the years, even if it includes properties that are closer to realistic situations, specially in biology, such as communication between oscillators taking place trough pulsed interactions and an explicit use of phase response curves \cite{smeal2010} in its formulation. It also displays a richer set of patterns, not only incoherence, partial locking and locking, as is seen in the Kuramoto model, but also mixed states and a regime of cessation of oscillation caused by excessively strong coupling, called \emph{oscillation death}.

Two notable exceptions to this absence of interest are the works of Ariaratnam and Strogatz in 2001, \cite{ariaratnam2001}, where the phase diagram of one version of the Winfree model was analyzed and of Pazó and Montibrió in 2014 \cite{pazo2014}, where the validity of the Ott-Antonsen ansatz \cite{ott2008,ott2009} for the Winfree model was demonstrated and used to obtain the time evolution of many important properties.

However, the previous works assumed a global pattern of connections, where each oscillator was equally coupled to all other. In this paper we would like to analyze the behavior of the Winfree model, specially the oscillation death regime, on complex networks, that, as far as we know, is absent in the literature. It's know that the topology of connections play a fundamental role in the synchronization process \cite{rodrigues2016} and therefore it's interesting to know how it affects the WM. Remarkably, we find that the oscillation death regime, where the oscillators become motionless, stationed at specific positions, follows very simple rules for a wide set of topologies. In fact, it shows universal properties that does not depend upon the topology of connections. To support our numerical experiments, we develop some analytical approximations, in the spirit of what was done for the Kuramoto model in \cite{pinto2015}, that follow closely the numerical values observed. 

This paper is organized as follows. In section (\ref{winfree_section}) we will discuss the essential facts about the Winfree model on complex networks. Later, on section (\ref{numerical_experiments}), we will perform our numerical experiments and develop a simple approximation to cover the results found. Section (\ref{degree_frequency_correlation}) treats the special case of degree-frequency correlation.

\section{The Winfree model on complex networks}
\label{winfree_section}

The Winfree model on a complex network of $N$ vertices is defined by the set of nonlinear differential equations

\begin{equation}
\dot{\varphi}_i = \omega_i + \lambda R(\varphi_i) \sum_{j=1}^{N} A_{ij} P_q(\varphi_j),
\label{winfree_model}
\end{equation}
where $\varphi_i$ and $\omega_i$, for $i=1,2,...,N$, are the phases and natural frequencies of the oscillators, $\lambda$ is the coupling strength and $A_{ij}$ denotes the adjacency matrix, with entry $A_{ij} = 1$ if vertices $i$ and $j$ are connected and $A_{ij} = 0$ otherwise. The natural frequencies are drawn from a symmetric and unimodal distribution $g(\omega)$ that we choose to be centered at $\omega = 1$ (note that model (\ref{winfree_model}) is not rotationally invariant as is the Kuramoto model, but by a suitable scaling of time we can always set $\langle \omega \rangle = 1$). This restriction is violated only in section \ref{degree_frequency_correlation}, where the positive degree-frequency correlation imposed on the natural frequencies may have a skewed distribution.  

In the limit of weak coupling and small diversity of natural frequencies, averaging theory can be used to show that the deviations from the mean phase $\varphi(t) = t$ follows the Kuramoto model \cite{pazo2014,kuramoto1984}.

The phase response curve $R(\varphi)$ determines how a oscillator responds to the impulses received from its neighbors and is chosen to be sinusoidal,

\begin{equation}
R(\varphi) = -\sin(\varphi),
\label{sensitivity_function}
\end{equation}
a form that qualitatively approximates many PRC measured in biological settings and is commonly employed \cite{ariaratnam2001,pazo2014}. The shape of the pulse is determined by the function 

\begin{equation}
P_q(\varphi) = a_{q} (1 + cos(\varphi))^{q}.
\label{influence_function}
\end{equation}

This function has a maximum at $\varphi = 0$, that is roughly when the oscillator fires, whose width can be controlled by the parametere $q$ such that the larger the $q$, the sharper the pulses and the normalization constant $a_q = 2^{q}(q!)^{2}/(2q)!$ is such that $\int_{0}^{2\pi} P_q(\varphi) d\varphi = 2\pi$.

\section{Numerical experiments}
\label{numerical_experiments}

It's know from \cite{ariaratnam2001,pazo2014} that for global coupling and for sufficiently large values of $\lambda$, a state of oscillation death emerges. In this case the systems approaches a fixed point where all the oscillations are motionless. We found numerically that on complex networks the oscillation death state still exists and the phases where the oscillators stay at rest can be very well approximated (for sufficiently large coupling strength $\lambda$) by the relations

\begin{equation}
\varphi_i = \begin{cases}
\alpha(q, \lambda) \omega_i /k_i   &\text{if $\omega_i > 0$}\\
2\pi - \alpha(q, \lambda) \omega_i /k_i &\text{if $\omega_i < 0$},
\end{cases}
\label{phases_asymptotic}
\end{equation}
where $k_i = \sum_{j=1}^{N} A_{ij}$ is the degree of the vertex $i$, its number of neighbors, and remarkably $\alpha$ is a function that depends only on $q$ and $\lambda$, it does not depend on the topology or other properties of the vertex in question. Even if relations (\ref{phases_asymptotic}) hold for positive and negative values of $\omega_i$, from now on we will only consider positive values, for ease of visualization and for simplicity in the calculations, even if all of our arguments can be trivially extended to this cases.

\begin{figure}
	\centering
	\includegraphics[width=0.7\linewidth]{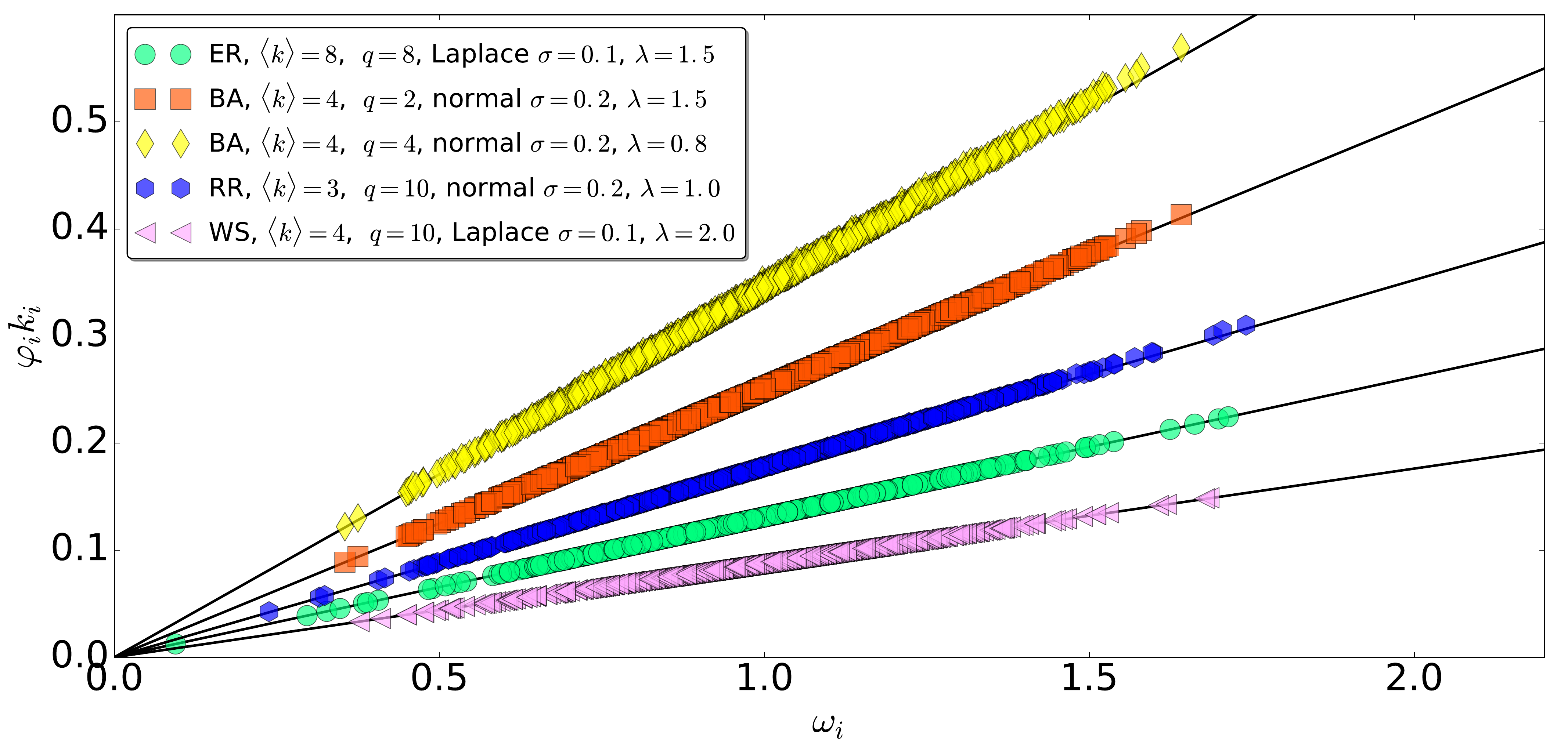}
	\caption{The asymptotic phases $\varphi_i$ times the degree $k_i$ of the vertex $i$ as a function of the natural frequency $\omega_i$ in the death state for different combinations of complex networks, natural frequency distributions and parameters $q$ and $\lambda$. ER, BA, RR and WS stand for Erdos-Renyi \cite{erdos1959}, Barabasi-Albert \cite{barabasi1999}, random regular and Watts-Strogatz \cite{watts1998} models, respectively. In particular, for the WS network, the rewiring probability $p = 0.2$. All distributions were centered at $\omega = 1.0$ and $\sigma$ stands for the standard deviation for normal distributions and the scale parameter for the Laplace distribution. All of the networks have $N = 2000$ vertices and different mean degree $\langle k \rangle$. The solid black line is the theoretical result (\ref{phases_asymptotic}) with $\alpha$ given by (\ref{alpha_value}).}
	\label{optimized_network_sync_diagram}
\end{figure}

Figure (\ref{optimized_network_sync_diagram}) shows that this scenery is true for different combination of networks and distributions $g(\omega)$, from the Barabasi-Albert network \cite{barabasi1999}, with its power law degree distribution, Erdos-Renyi network \cite{erdos1959}, possessing a Poissonian degree distribution, random regular networks, where all vertices have the same degree but the connections are random and even to Watts-Strogatz networks \cite{watts1998}, that are highly clustered networks.

Up to now, we only found that the phases follows the relations (\ref{phases_asymptotic}), but the precise dependence of the function $\alpha$ on the parameters $q$ and $\lambda$ is still missing. To fill this gap, we can try an approximation inspired in what was done for the Kuramoto model in \cite{pinto2015}. Let's assume that (\ref{phases_asymptotic}) is a fixed point of the system. Plugging it in (\ref{winfree_model}), with the condition $\dot{\varphi}_{i} = 0$ and summing with respect to the index $i$, $\alpha$ can be found as the root of the following equation

\begin{equation}
f(\alpha) = \sum_{i=1}^{N} \omega_i - \lambda \sum_{i=1}^{N} \sum_{j=1}^{N} A_{ij}\sin(\alpha \omega_i /k_i) P_{q}(\alpha \omega_j /k_j).
\label{alpha_equation}
\end{equation}

\begin{figure}
	\centering
	\includegraphics[width=0.7\linewidth]{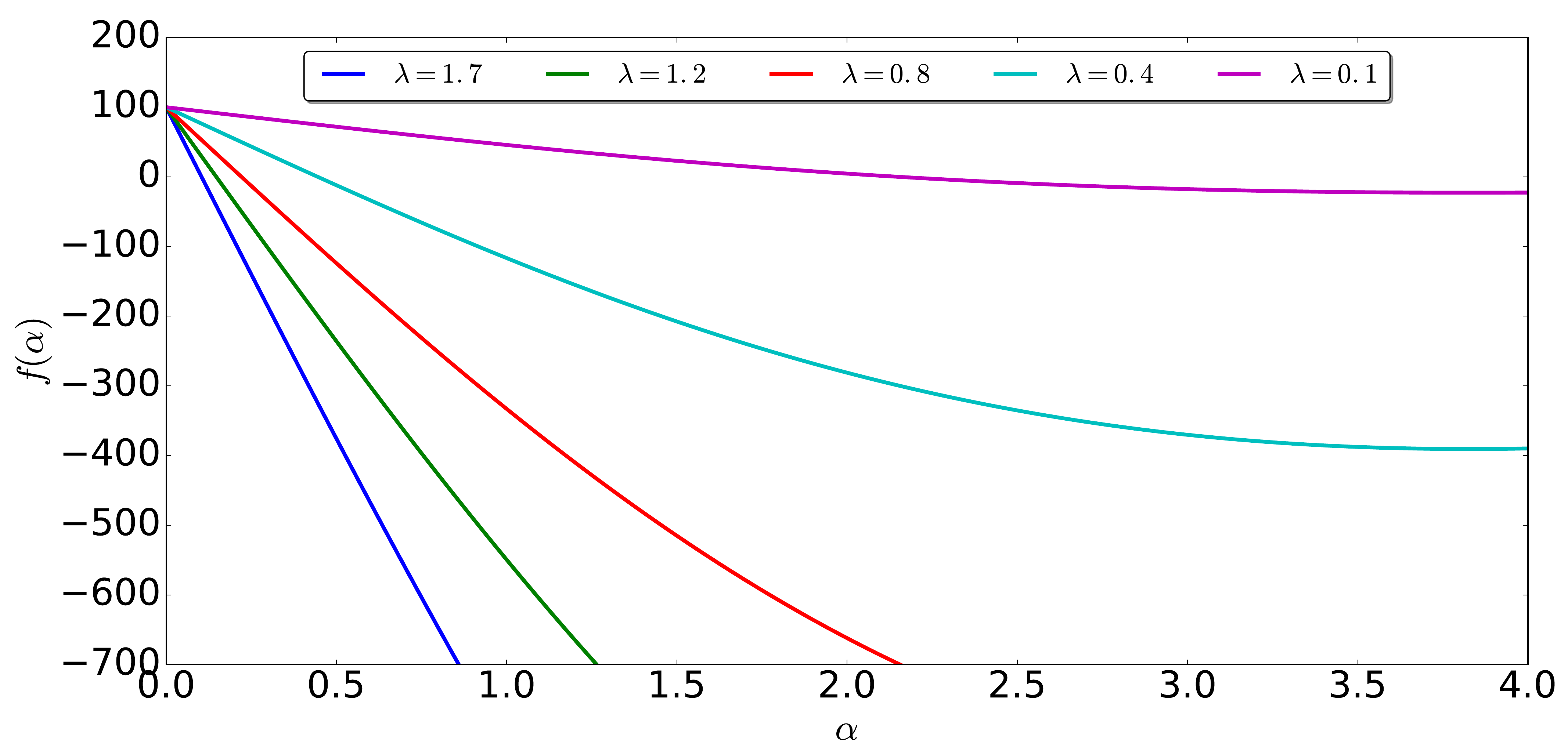}
	\caption{The function $f(\alpha)$, (\ref{alpha_equation}), for an Erdos-Renyi network of $N=100$ vertices, natural frequencies drawn from a normal distribution, $q = 10$ and different values of $\lambda$.}
	\label{f_behavior}
\end{figure}

Even if solving (\ref{alpha_equation}) analytically is probably impossible, we can do it approximately. By numerical inspection, see Figure \ref{f_behavior}, we find that for a large range of $\alpha$ values, starting at  $\alpha = 0$, the behavior of the function $f(\alpha)$ is almost perfectly linear. Moreover, the root we are interested lies precisely in this range. Therefore we can linearize $f(\alpha)$ at $\alpha = 0$, and then find where it crosses the horizontal axis, giving us the value

\begin{equation}
\alpha(q, \lambda) = \frac{1}{2^{q} a_{q} \lambda}.
\label{alpha_value}
\end{equation}

In accordance with the numerical experiments performed, the function $\alpha$ we found depends only on the width of the pulses, trough the parameter $q$, and on the coupling strength $\lambda$, such that $\alpha$ decreases as $q$ increases. It's to be noted the goodness of the approximation, as (\ref{phases_asymptotic}) with $\alpha$ given by (\ref{alpha_value}) is very close to the numerical values (see Figure \ref{optimized_network_sync_diagram}). 

\begin{figure}
	\centering
	\includegraphics[width=0.7\linewidth]{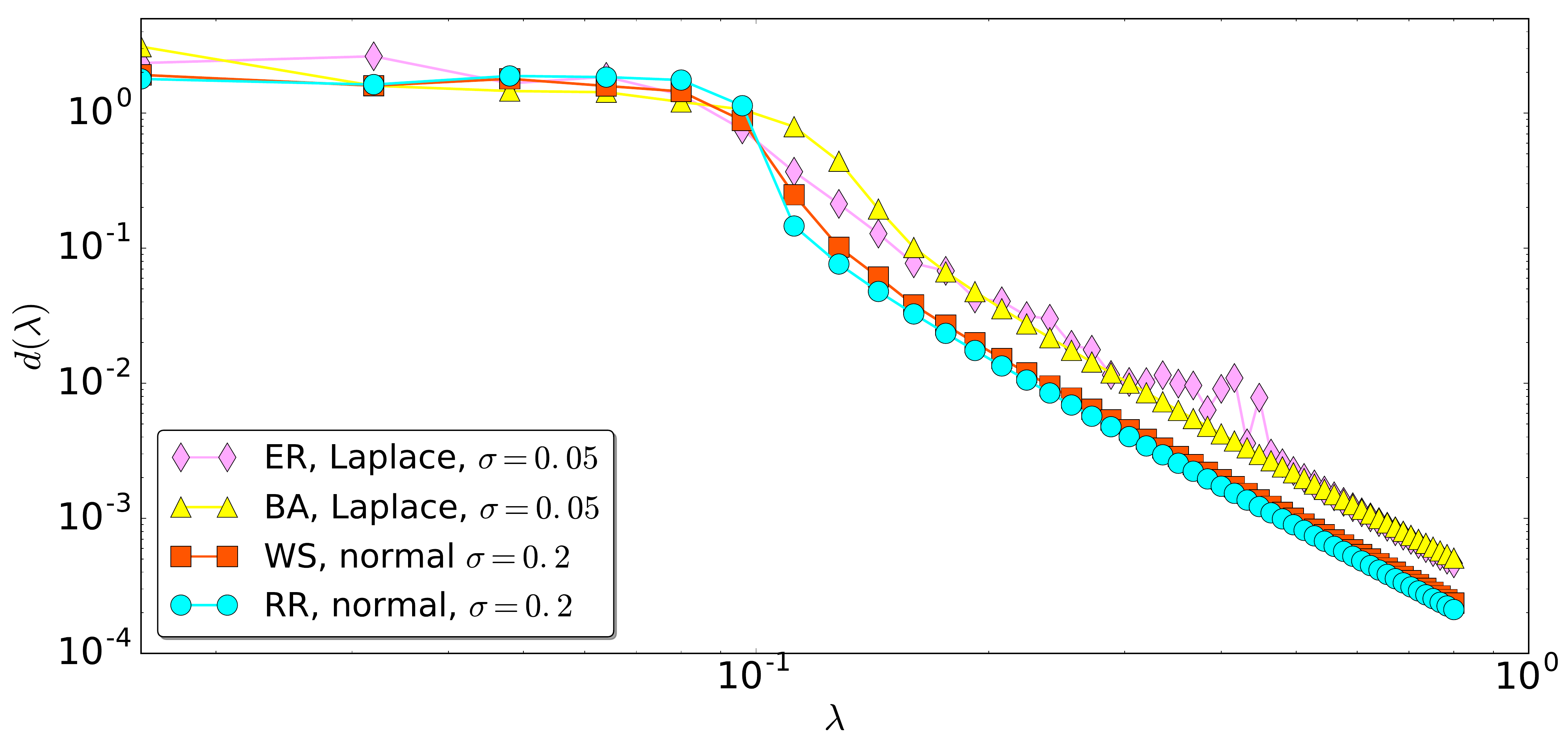}
	\caption{The distance (\ref{definition_d}) between the current state (as a function of $\lambda$) to the oscillation death regime described by (\ref{phases_asymptotic}) for the networks used in Figure (\ref{optimized_network_sync_diagram}). $q = 10$ and different combinations of networks and distributions $g(\omega)$ were used, as indicated in the legend.}
	\label{decay_death_state}
\end{figure}

Approximation (\ref{phases_asymptotic}) improves its ability to model the numerical results with increasing coupling strength. In order to quantify how fast solutions of (\ref{winfree_model}), with appropriated parameters, approach, as a function of the coupling strength, the state of oscillation death described by (\ref{phases_asymptotic}), we introduce the distance function 

\begin{equation}
d(\lambda) = \frac{1}{N} \sum_{i=1}^{N} \left| \varphi_i - \alpha\omega_i / k_i \right|,
\label{definition_d}
\end{equation}
and apply it for the networks used as example in Figure (\ref{optimized_network_sync_diagram}). Our results, depicted in Figure \ref{decay_death_state}, show us that $d(\lambda)$ first has an almost constant value, that, due to the parameters used, correspond to the incoherent state. From  $\lambda \approx 0.1$ onwards, oscillators start to die and $d(\lambda)$ changes its behavior and decays as a power law. Moreover, the decay rate depends much more stronger on the frequencies drawn than on the topology employed, as one can see in Figure \ref{decay_death_state}, where different networks, but with the same frequencies, shows almost the same decay.

\section{The case of degree-frequency correlation}
\label{degree_frequency_correlation}

Finally, we would like to analyze what happens to oscillation death in the case of degree-frequency correlation, a situation that attracted a lot of attention for the Kuramoto model \cite{boccaletti2016}. In this case, we will take $\omega_i = b k_i$, with $b > 0$. For this particular situation, the phase's form (\ref{phases_asymptotic}) still hold true, but now $\alpha$ is a function of $b$ and $\lambda$ (actually, all the phases have the same value in this case, $\varphi_i = b \alpha(b, \lambda) /\lambda$). The interest in this situation is that equation (\ref{alpha_equation}) simplifies tremendously,

\begin{equation}
f(\alpha) = 2m \left( b - \lambda \sin(b\alpha)P_{q}(b\alpha) \right),
\label{equation_alpha_df}
\end{equation}
where $m = 1/2 \sum_{i=1}^{N}\sum_{j=1}^{N} A_{ij}$ is the number of edges in the network. The term $\sin(b\alpha)P_{q}(b\alpha)$ is oscillatory but has a maximum value that does not depend upon $b$, as it only plays the role of frequency in (\ref{equation_alpha_df}). In order that (\ref{equation_alpha_df}) has a root, this term must, for some values of $\alpha$ be larger than $b/\lambda$. Therefore, by choosing progressively larger values of $b$, the necessary value of the coupling strength such that (\ref{equation_alpha_df}) has a root, and so produce oscillation death, is pushed to larger and larges values. 

\begin{figure}
	\centering
	\includegraphics[scale=0.2]{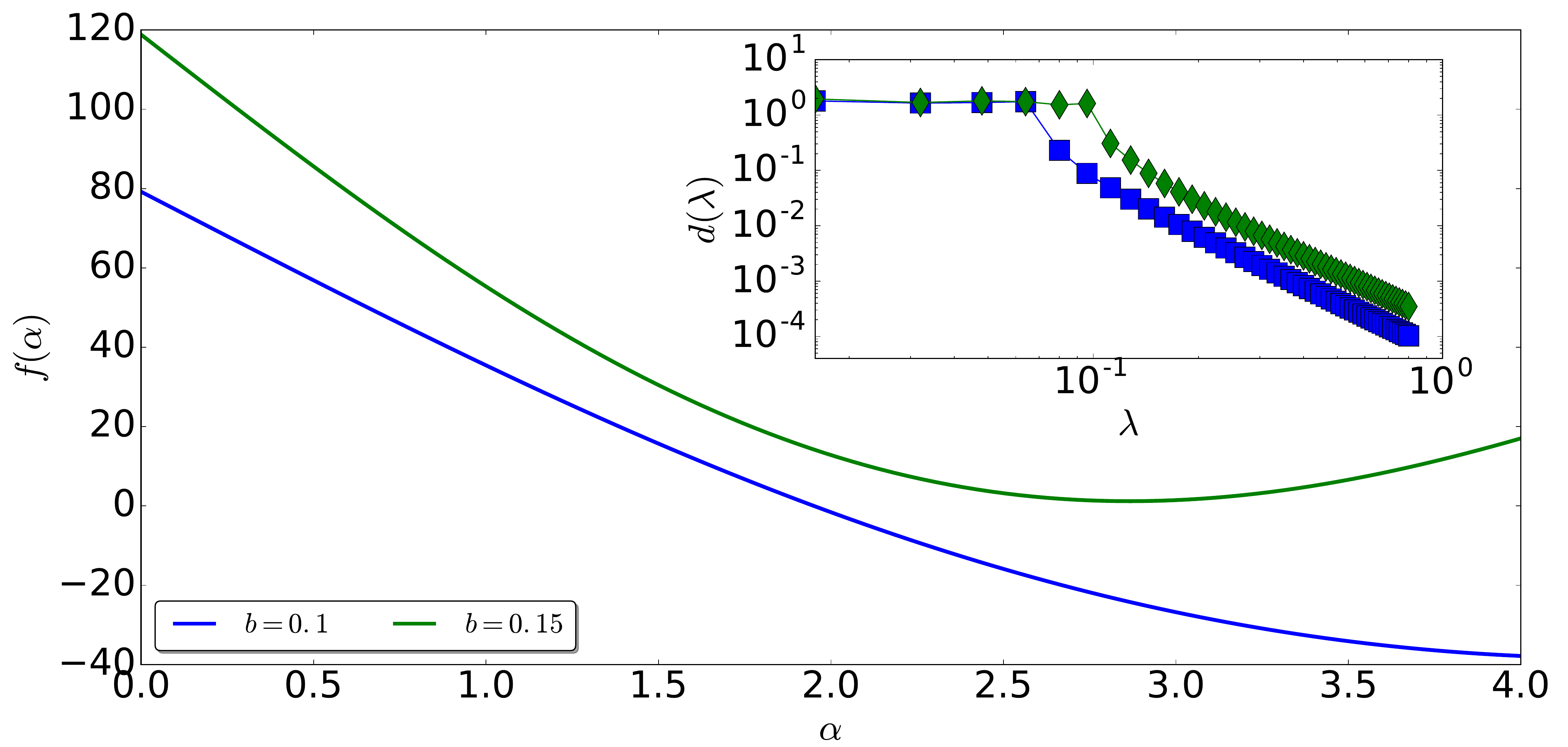}
	\caption{This figure shows the function $f(\alpha)$ for the case of degree-frequency correlation, equation (\ref{equation_alpha_df}), for $\lambda = 0.1$ and different values of the parameter $b$ (green lines and diamonds for $b = 0.15$ and blue lines and squares for $b = 0.1$). The inset shows the distance function (\ref{definition_d}) for both cases. In the calculations, $q = 1$.}
	\label{correlation}
\end{figure}

An example of this is depicted in Figure \ref{correlation}, for an Erdos-Renyi network, where one can see the effect of increasing $b$ in the appearance of oscillation death. Whereas we have that $f(\alpha)$ has a root at parameters $\lambda = 0.1$ and $\beta = 0.1$, and so oscillation death, when we keep $\lambda = 0.1$, but increase $\beta = 0.15$, the function $f(\alpha)$ change its behavior and does not have a root anymore (and therefore, oscillation death). This reflects in the diagram of the distance (\ref{definition_d}) as a function of $\lambda$, shown in the inset, where $d$ is large for $\lambda = 0.1$ when $b = 0.15$, but already entered the region of decay for $b = 0.1$.

\section{Conclusions}

We have analyzed the oscillation death state for the phase model proposed by Winfree \cite{winfree1967} and found that the motionless oscillators stay are rest in positions that depends on the ratio of their natural frequency to their degree times a constant that depends only on the coupling strength $\lambda$ and on the parameter $q$ that controls the width of the pulses. Beyond the appealing simplicity of (\ref{phases_asymptotic}), it also shows that two different networks that have the same frequencies and degrees, but are otherwise different (different clustering, associativity or any other property) have that their phases freeze at the same exactly values. Furthermore, in the case of degree-frequency oscillation, we can show that it inhibits the emergence of oscillation death.

We also found that property (\ref{phases_asymptotic}) also holds for more general PRCs, such as $R(\varphi) = \sin(\beta) - \sin(\beta + \varphi)$, \cite{pazo2014}, for $0 \le \beta < \pi/2$, where the only difference is that now equation (\ref{alpha_value}) reads $\alpha = \left( \lambda \cos(\beta) a_q 2^q \right)^{-1}$.

The results presented here complement a similar situation (albeit from a different origin) where a ensemble of coupled oscillators goes to a steady state is \emph{amplitude death} \cite{saxena2012}. This situation is generally observed in coupled amplitude (Stuart-Landau) oscillators, where a unstable fixed point becomes stable when the coupling among oscillators is strong enough and if some conditions (such as frequency mismatch) are valid. 

Finally, simple relations such as the ones found for the oscillation death state cannot be extended to the synchronization case, inhibiting the use of approximations like the ones studied in \cite{gottwald2015} and methods like those in \cite{pinto2015} for optimizing the network either to favor or to inhibit oscillation death. Nevertheless, by sorting with caution the natural frequencies, one can target death states with more or less heterogeneity (with respect to the distribution of the phases).

\section*{Acknowledgements}

The author thank CNPq for the financial support. Our numerical computations were done by using the SciPy package for python \cite{jones2001} and the NetworkX package \cite{hagberg2008}.


\begin{thebibliography}{99}

\bibitem{winfree1967}
A. T. Winfree, J. Theor. Biol. 15, 16 (1967).

\bibitem{winfree1980}
A.T. Winfree, The Geometry of Biological Time, (Springer, New York, 1980).

\bibitem{strogatz2000}
S. H. Strogatz, Physica D {\bf 143}, 1 (2000).

\bibitem{kuramoto1975}
Y. Kuramoto, in {\em Proceedings of the International Symposium on Mathematical Problems in Theoretical Physics}, 
University of Kyoto, Japan, Lect. Notes in Physics {\bf 30}, 420  (1975), edited by H. Araki.

\bibitem{acebron2005}
J. A. Acebrón, L. L. Bonilla, C. J. P. Vicente, F. Ritort, and R. Spigler, Rev. Mod. Phys. {\bf 77}, 137 (2005).

\bibitem{rodrigues2016}
F. A. Rodrigues, T. K. D. Peron, P. Ji, and J. Kurths, Phys. Rep. 610, 1 (2016).

\bibitem{smeal2010}
R. M. Smeal, G. B. Ermentrout, and J. A. White, Phil. Trans. R. Soc. B 365, 2407 (2010).

\bibitem{ariaratnam2001}
J. T. Ariaratnam, and S. H. Strogatz, Rev. Lett. 86, 4278 (2001).

\bibitem{pazo2014}
D. Pazó, and E. Montbrió, Phys. Rev. X 4, 011009 (2014).

\bibitem{ott2008}
E. Ott, and T. M. Antonsen, Chaos {\bf 18}, 037113 (2008).

\bibitem{ott2009}
E. Ott, and T. M. Antonsen, Chaos {\bf 19}, 023117 (2009).

\bibitem{pinto2015}
R. S. Pinto, and A. Saa, Phys. Rev. E {\bf 92}, 062801 (2015).

\bibitem{kuramoto1984}
Y. Kuramoto, Chemical Oscillations, Waves, and Turbulence (Springer-Verlag, Berlin, 1984).

\bibitem{barabasi1999}
A.~L.~Barabasi, and R.~Albert, Science {\bf 286}, 5439 (1999).

\bibitem{erdos1959}
P. Erdos, and A. Renyi, Publ. Math. Inst. Hung. Acad. Sci. 5, 17 (1960).

\bibitem{watts1998}
D. J. Watts, and S. H. Strogatz, Nature, 393, (1998).

\bibitem{boccaletti2016}
S. Boccaletti, J. A. Almendral, S. Guan, I. Leyva, Z. Liu, I. Sendiña-Nadal, Z. Wang, and Y. Zou, \href{https://arxiv.org/abs/1610.01361}{https://arxiv.org/abs/1610.01361} (2016).

\bibitem{saxena2012}
G. Saxena, A. Prasad, and R. Ramaswamy, Phys. Rep. 521, 205 (2012).

\bibitem{gottwald2015}
G. Gotwald, Chaos {\bf 25}, 053111 (2015).

\bibitem{jones2001}
E. Jones, E. Oliphant, P. Peterson, et al., SciPy: Open Source Scientific Tools for Python (2001), \href{http://www.scipy.org}{http://www.scipy.org} 

\bibitem{hagberg2008}
A. A. Hagberg, D. A. Schult, and P. J. Swart, Exploring network structure, dynamics, and function using NetworkX, in Proceedings of the 7th Python in Science Conference, edited by
G. Varoquaux, T. Vaught, and J. Millman (Pasadena, CA, 2008), pp. 11-15.


\end{thebibliography}
\end{document}